\documentclass[twocolumn,aps,prl,showpacs]{revtex4}

\usepackage{epsfig}
\usepackage{graphicx}
\usepackage{amsmath}
\usepackage{bm}

\begin{document}

\title{Identifying Collective Modes via Impurities in the Cuprate Superconductors}
\author{Roy H. Nyberg, Enrico Rossi and Dirk K. Morr}
\affiliation{Department of Physics, University of Illinois at
Chicago, Chicago, IL 60607}
\date{\today}
\begin{abstract}

We show that the pinning of collective charge and spin modes by
impurities in the cuprate superconductors leads to qualitatively
different fingerprints in the local density of states (LDOS). In
particular, in a pinned (static) spin droplet, the creation of a
resonant impurity state is suppressed, the spin-resolved LDOS
exhibits a characteristic spatial pattern, and the LDOS undergoes
significant changes with increasing magnetic field. Since all of
these fingerprints are absent in a charge droplet, impurities are a
new probe for identifying the nature and relative strength of
collective modes.

\end{abstract}


% 74.25.Ha    Magnetic properties
% 74.25.Jb    Electronic structure
% 73.20.Mf    Collective excitations
% 74.72.-h    Cuprate superconductors

\maketitle

Whether the unconventional properties of the high-temperature
superconductors (HTSC) arise from the interaction of electronic
degrees of freedom with collective spin modes, charge modes, or
phonons, is one of the key issues in understanding these complex
materials \cite{reviews}. While some angle-resolved photoemission
(ARPES) experiments \cite{Campuzano:1999} observed effects of the
magnetic {\it resonance mode} \cite{INS,Tranquada:2005} on
electronic excitations, other ARPES studies argued that the
electronic dispersion exhibits a number of prominent features that
arise from the coupling to phonons \cite{Lanzara}. Moreover, recent
scanning tunneling spectroscopy (STS) experiments \cite{Lee:2006}
have reported evidence for a collective mode whose frequency is
doping independent and shifts with isotope substitution, two
properties that are inconsistent with those of the magnetic
resonance mode \cite{Tranquada:2005,Pailhes:2005}. The
interpretation of these experiments is further complicated by the
coupling between charge modes and phonons \cite{Tranquada:2005,
Reznik:2006}. Clearly, new experiments are required that can
unambiguously identify the most relevant collective mode, and its
effect on electronic excitations in the HTSC.

In this Letter, we propose that STS experiments can directly address
this issue by studying the effects in the LDOS of a
$d_{x^2-y^2}$-wave superconductor that arise from a coupling between
impurities and collective spin and charge modes. In particular, we
show that such a coupling induces a static spin (charge) density
droplet which represents an extended scattering potential for the
electronic degrees of freedom and thereby affects the
superconductor's local electronic structure. Indeed, static spin
droplets around Ni and Zn impurities have been observed in nuclear
magnetic resonance (NMR) experiments
\cite{Bobroff:1997,Julien:2000,Ouazi:2004,Ouazi:2006}. By using the
${\hat T}$-matrix \cite{Shiba68,Sta06} and Bogoliubov-de Gennes
(BdG) \cite{Gen89} formalisms, we demonstrate that scattering off
spin and charge droplets gives rise to {\it qualitatively} different
fingerprints in the superconductor's LDOS. In particular, we find
that the coupling of an impurity to a spin mode peaked at ${\bf
Q}=(\pi,\pi)$ prevents the creation of a resonant impurity state,
leads to characteristic spatial patterns of the spin-resolved LDOS
directly reflecting the mode's antiferromagnetic nature, and gives
rise to a pronounced magnetic field dependence of the LDOS. Since
all of these features are absent in the presence of a charge
droplet, our results demonstrate that impurities can be used to
identify the nature and relative strength of collective modes not
only in the HTSC, but in strongly correlated electron systems in
general.

The coupling of a single impurity with spin ${\bf S}_{imp}$ located
at site $\mathbf{R}$ to collective spin and charge modes,
represented by the operators $\mathbf{s}_e$ and $n_e$, respectively,
is described by the Hamiltonian \cite{Morr:1998b,DellAnna:2005}
\begin{equation}
 \mathcal{H}_{int}=-J \mathbf{S}_{imp} \cdot \mathbf{s}_e({\bf R}) + U n_{imp}\, n_e({\bf R})
 \label{eq:ham_imp_int}
\end{equation}
with $J, U>0$ and $\langle n_{imp} \rangle=1$. Note that the
creation of a static spin droplet requires a non-zero spin
polarization of the impurity, $\langle S^z_{imp} \rangle$, which can
be induced, for example, by applying a magnetic field $H \ll
H^{ab}_{c2}$ in the $ab$-plane thus avoiding complications arising
from the creation of vortices \cite{com4}. Experimentally, $\langle
S^z_{imp} \rangle =C H/(T+\Theta)$ obeys the Curie-Weiss law
\cite{Ouazi:2006}. The impurity-mode coupling of
Eq.(\ref{eq:ham_imp_int}) induces static spin \cite{Morr:1998b} and
charge density oscillations \cite{DellAnna:2005} described by
\begin{eqnarray}
 \langle s_e^z (\mathbf{r}) \rangle &=& J \langle S^z_{imp} \rangle
 \chi_s( \mathbf{r}-\mathbf{R},\omega=0)  \
 \ , \nonumber \\
\langle \delta n_e (\mathbf{r}) \rangle &=&  U  \langle n_{imp}
\rangle \chi_c(\mathbf{r}-\mathbf{R},\omega=0) \ ,
\end{eqnarray}
respectively. Here, $\chi_s (\chi_c)$ is the spin (charge)
susceptibility, $\delta n_e (\mathbf{r})=n_e (\mathbf{r})-n_0$, and
$n_0$ is the uniform charge density. For the static susceptibilities
in momentum space, we make the ansatz $\chi_{s,c}({\bf
q},\omega=0)=\chi^{s,c}_0/(\xi_{s,c}^{-2}+({\bf q}-{\bf
Q}_{s,c})^2)$ where $\xi_{s,c}$ is the respective correlation
length, ${\bf Q}_s=(\pi,\pi)$ \cite{Scalapino:1995} and hence ${\bf
Q}_c = 0$, in agreement with NMR experiments
\cite{Bobroff:1997,Julien:2000,Ouazi:2004,Ouazi:2006,Morr:1998b,Harter:2006}.
The mean-field Hamiltonian of the entire system is given by
\begin{eqnarray}
 \mathcal{H}&=&\sum_{\mathbf{r,r'},\sigma} t_{\bf rr'}
 c^\dagger_{\mathbf{r} ,\sigma} c_{\mathbf{r'},\sigma}
 +\sum_{\mathbf{r,r'}} \left[ \Delta_{\bf r,r'} c^\dagger_{
 \mathbf{r},\uparrow} c^\dagger_{\mathbf{r'},\downarrow} + h.c.
 \right] \nonumber \\
 & & \hspace{-0.5cm} -\sum_{\mathbf{r},\alpha,\beta}
 \left[
       g_s\langle s_z(\mathbf{r}) \rangle \sigma^z_{\alpha \beta} -
       g_c\langle \delta n (\mathbf{r}) \rangle {\bf 1}_{\alpha \beta}
 \right]
 c^\dagger_{\mathbf{r},\alpha} c_{\mathbf{r},\beta} \ ,
 \label{H}
\end{eqnarray}
where $c^{\dagger}_{\mathbf{r},\sigma}$ creates an electron with
spin $\sigma$ at site ${\bf r}$, $t_{\bf rr'}$ is the hopping
integral between sites ${\bf r}$ and ${\bf r'}$, and $\Delta_{\bf
r,r'}$ is the $d_{x^2-y^2}$-wave superconducting (SC) gap. The last
term in Eq.(\ref{H}) describes the scattering of electrons by the
total spin density $\langle s_z(\mathbf{r}) \rangle = \langle s_e^z
(\mathbf{r}) \rangle + \langle S^z_{imp} \rangle
\delta_{\mathbf{r},\mathbf{R}}$ and effective charge density
 $\langle \delta n (\mathbf{r}) \rangle = \langle \delta n_e
(\mathbf{r}) \rangle + \delta_{\mathbf{r},\mathbf{R}} \langle
n_{imp} \rangle$. The droplets' scattering strength is determined by
only two parameters: for a spin droplet by $\eta_s=J \chi_s(0,0)$
and ${\bar g}_s = g_s \langle S^z_{imp} \rangle$, such that
$g_s\langle s_z(\mathbf{r}) \rangle ={\bar g}_s\left[\eta_s
\chi_s({\bf r-R},0)/\chi_s(0,0)+\delta_{\bf R,r}\right]$, and for a
charge droplet by $\eta_c=U \chi_c(0,0)$ and ${\bar g}_c=g_c \langle
n_{imp} \rangle$.

We study the effects of electronic scattering on the LDOS by using
two complementary methods: the ${\hat T}$-matrix
\cite{Shiba68,Sta06} approach, which allows us to investigate large
host systems but assumes a spatially constant SC order parameter
(SCOP), and the Bogoliubov-de Gennes (BdG) \cite{Gen89} formalism,
which accounts for spatial variations of the SCOP, but can only
treat small system sizes. Within the ${\hat T}$-matrix approach, the
Green's function matrix in Matsubara space is given by \cite{Sta06}
\begin{eqnarray}
 \hat{G}(\mathbf{r},\mathbf{r^{\prime}},\omega_n)&=&\hat{G}_0(\mathbf{r},
 \mathbf{r^{\prime}},\omega_n)  \nonumber  \\
 & & \hspace{-2cm} +\sum_{\bf l,p}
 \hat{G}_0(\mathbf{r},\mathbf{l},\omega_n)
 \hat{T}(\mathbf{l},\mathbf{p},\omega_n)\hat{G}_0(\mathbf{p},
 \mathbf{r^{\prime}},\omega_n) \ ,  \label{Ghat}
 \end{eqnarray}
where the sum runs over all droplet sites. The ${\hat T}$-matrix is
determined from
\begin{equation}
\hat{T}(\mathbf{l},\mathbf{p},\omega_n)= \hat{V}_{\bf l} \delta_{\bf
l,p} +\hat{V}_{\bf l} \sum_{\bf s}
\hat{G}_0(\mathbf{l},\mathbf{s},\omega_n)\hat{T}(\mathbf{s},\mathbf{p},\omega_n)
\ ,
\end{equation}
where
$\hat{G}_0= \left[ i\omega_n \sigma_0 - \varepsilon_{%
\mathbf{k}} \sigma_3 + \Delta_{\mathbf{k}} \sigma_1 \right]^{-1}$ is
the unperturbed Green's function, ${\hat V}_{\bf l}=-g_s \langle
s_z(\mathbf{l}) \rangle \sigma^z + g_c\langle \delta n (\mathbf{l})
\rangle {\bf 1} $, $\varepsilon_{\mathbf{k}}=-2t \left( \cos k_x +
\cos k_y \right) -4t' \cos k_x \cos k_y-\mu$ is the normal state
tight binding dispersion with $t=300$ meV, $t'/t=-0.4$, and
$\mu/t=-1.083$, representative of the HTSC \cite{ARPES}, and
$\Delta_k = \Delta_{\bf 0} (\cos{k_{\bf x}} - \cos{k_{\bf y})}/2$ is
the SC gap with $\Delta_0=30$ meV.  The LDOS,
$N(\mathbf{r},\omega)=A_{11}(\mathbf{r},\omega)-A_{22}(\mathbf{r},-\omega)$
with $A_{ii}(\mathbf{r},\omega)= -\mathrm{Im}\,
\hat{G}_{ii}(\mathbf{r},\omega+i\delta)/ \pi$ and $\delta=0.2$ meV
is obtained from Eq.(\ref{Ghat}). In contrast, in the BdG formalism
\cite{Gen89}, one solves the eigenvalue equation
\begin{eqnarray}
& & \sum_{\bf r'}
\begin{pmatrix}
 H^+_{\bf rr'} & \Delta_{\bf rr'} \\
\Delta^*_{\bf rr'} & -H^-_{\bf rr'}
\end{pmatrix}
\begin{pmatrix}
u_{{\bf r'},n} \\
v_{{\bf r'},n}
\end{pmatrix}
= E_n
\begin{pmatrix}
u_{{\bf r},n} \\
v_{{\bf r},n}
\end{pmatrix}
\label{EVeq}
\end{eqnarray}
with $H^\pm_{\bf r r'}=t_{\bf rr'}+(\mp g_s \langle
s_z(\mathbf{r})\rangle
 - g_c\langle n(\mathbf{r})\rangle-\mu) \delta_{\bf r,r'}$, and
 self-consistently computes the SC gap via
\begin{equation}
\Delta_{\bf rr'}=-\frac{V}{2} \sum_n \left[ u_{n}({\bf r})
v_{n}({\bf r'}) +u_{n}({\bf r'}) v_{n}({\bf r}) \right] \tanh\left(
\frac{E_n}{2 k_B T} \right) \ , \label{eq:BdGgap}
\end{equation}
where the sum runs over all eigenstates of the system, and
$V=0.7375\;t$ yields the same (clean) SC gap as taken in the ${\hat
T}$-matrix approach. The LDOS is obtained via
$$
N(\omega, {\bf r})=\sum_n \left[ u^2_{n}({\bf r}) \delta(\omega-E_n)
+ v^2_{n}({\bf r}) \delta(\omega+E_n) \right] .
$$

\begin{figure}[h]
 \epsfig{file=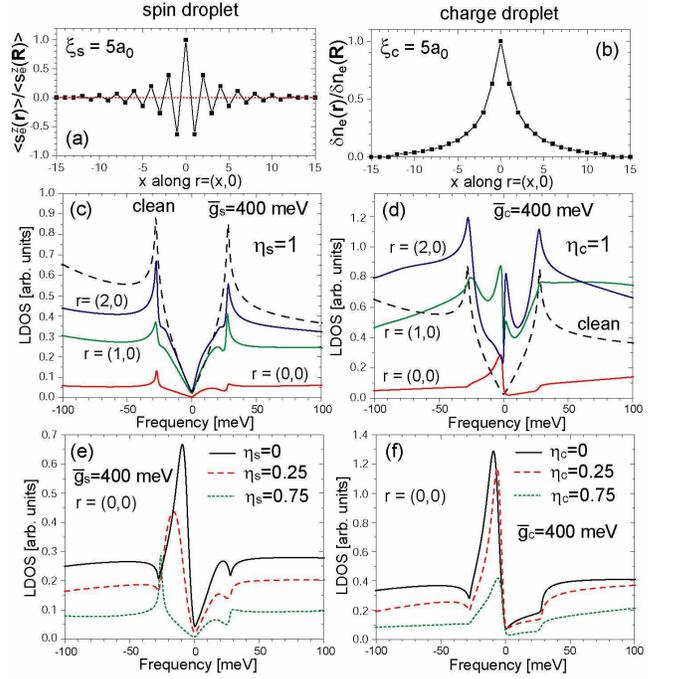,width=8.5cm}
 \caption{(color online) Normalized spin (a) and charge (b) density
along ${\bf r}=(x,0)$. LDOS inside a spin (c) and charge (d) droplet
with $\eta_{s,c}=1$ and ${\bar g}_s, {\bar g}_c=400$ meV (the dashed
lines represent the clean LDOS). Evolution of the LDOS with
increasing $\eta_{s,c}$ for a spin (e) and charge (f) droplet.}
 \label{fig1}
\end{figure}

In order to identify the qualitative effects of spin and charge
droplets on the LDOS, we first consider the case when only one of
the modes couples to the impurity and present in Fig.~\ref{fig1}(a)
[(b)] the normalized spin (charge) density for a pure spin (charge)
droplet [the center of the droplet is located at $(0,0)$]
\cite{com3}.
%
%  Fig.1
%
To directly compare the effects of the droplets, we use ${\bar
g}_s={\bar g}_c$ and $\xi_{c}=\xi_s$ with $\xi_s=5 a_0$
representative of the underdoped HTSC \cite{Ouazi:2004} and present
in Figs.~\ref{fig1} (c) and (d) the LDOS obtained from the ${\hat
T}$-matrix approach inside the spin and charge droplet,
respectively. The low-frequency LDOS in these two droplets exhibits
significant qualitative differences. Inside the spin droplet, the SC
coherence peaks are clearly visible and no impurity resonance exists
inside the SC gap (the LDOS exhibits, however, weak Friedel-like
oscillations). In contrast, inside the charge droplet, the SC
coherence peaks are strongly suppressed and a resonant impurity
state is formed with peaks in the LDOS at $\pm 2$ meV. These
qualitatively different effects of the charge and spin droplet on
the LDOS arise from the spatial forms of their scattering
potentials. For a spin droplet, the alternating sign of $\langle
s_z(\mathbf{r}) \rangle $ and hence of the scattering potential
leads to destructive interference of scattered electrons which
prevents the creation of an impurity state inside the SC gap. This
interpretation is supported by the fact that the impurity resonance
of the decoupled ($\eta_s=0$) impurity [see Fig.~\ref{fig1}(e)]
shifts to higher energies with increasing $\eta_s$, implying a
decrease in the effective scattering strength of the impurity. Note
that for sufficiently large $\eta_s$, the LDOS is qualitatively
different from that of a decoupled impurity ($\eta_s = 0$). In
contrast, for the charge droplet, the scattered electrons interfere
constructively due to the same sign of the scattering potential at
all sites, which leads to an increase in the effective scattering
strength. Accordingly, the impurity resonance of the decoupled
($\eta_c=0$) impurity shifts to lower energies with increasing
$\eta_c$ [see Fig.~\ref{fig1}(f)]. The modes' different momentum
dependence thus leads to distinct fingerprints of a spin
[Fig.~\ref{fig1}(c)] and charge [Fig.~\ref{fig1}(d)] droplet in the
LDOS.

\begin{figure}[h]
 \epsfig{file=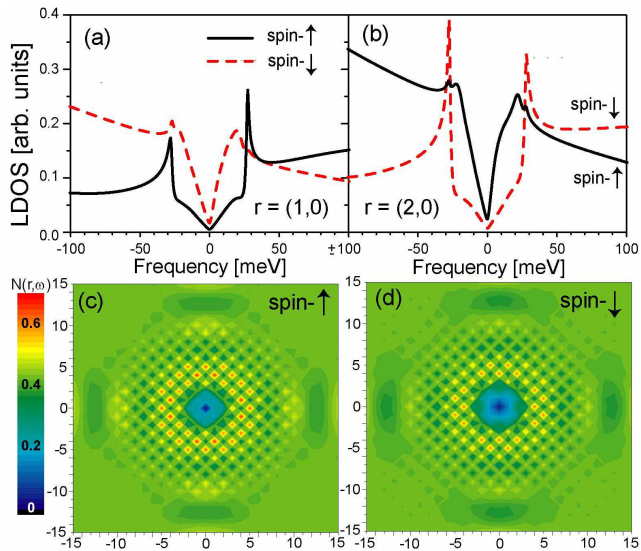,width=8.5cm}
\caption{(color online) Spin resolved LDOS for ${\bar g_s}=400$ meV
at adjacent sites (a) ${\bf r}=(1,0)$, and (b) ${\bf r}=(2,0)$.
Intensity plot of the (c) spin-$\uparrow$ and (d) spin-$\downarrow$
LDOS at $\omega_{cp}=30$ meV.}
 \label{fig2}
\end{figure}

Another important fingerprint of the spin droplet can be found in
the spatial structure of the spin-resolved LDOS, as shown in
Fig.~\ref{fig2}(a) and (b) where we present the spin-$\uparrow$ and
spin-$\downarrow$ LDOS at adjacent sites.
%
%  Fig.2
%
As expected from the spatially alternating sign of the scattering
potential, the frequency dependence of the spin-$\uparrow$ LDOS at
${\bf r}=(1,0)$ is qualitatively similar to that of the
spin-$\downarrow$ LDOS at ${\bf r}=(2,0)$, and vice versa. This
behavior is observed for all sites of the droplet that belong to
different (antiferromagnetic) sublattices. As a result, the
spin-$\uparrow$ and spin-$\downarrow$ LDOS possess a spatially
complementary intensity pattern, as shown in Figs.~\ref{fig2}(c) and
(d) where we present an intensity plot of the spin-resolved LDOS at
the frequency of the hole-like coherence peak, $\omega_{cp}=30$ meV.
At sites where the spin-$\uparrow$ LDOS is large, the
spin-$\downarrow$ LDOS is small, and vice versa.  In contrast, in a
charge droplet, the spin-$\uparrow$ and spin-$\downarrow$ LDOS are
identical. Hence, the qualitatively different behavior of the
spin-resolved LDOS in spin and charge droplets is directly linked to
the magnetic/non-magnetic nature of the modes.

\begin{figure}[h]
\epsfig{file=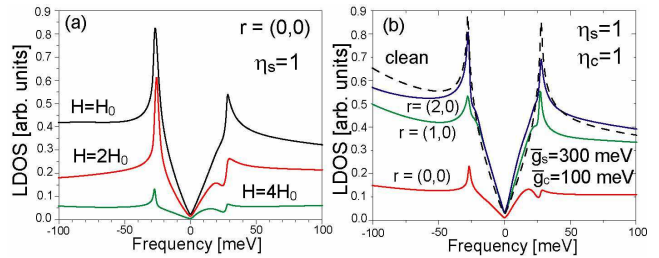,width=8.5cm} \caption{(color online) (a) Total
LDOS for a spin droplet at ${\bf r}=(0,0)$ for several ${\bar
g_s}=g_s \langle S^z_{imp}(H,T) \rangle$ ($U=0$). (b) Total LDOS for
a droplet with $\langle s_z(\mathbf{r}) \rangle, \langle \delta n
(\mathbf{r}) \rangle \not = 0$ and $\alpha=1/3$.} \label{fig3}
\end{figure}

Another qualitative difference between spin and charge droplets
arises from the magnetic field dependence of the impurity's spin
polarization $\langle S^z_{imp}(H,T) \rangle =C H/(T+\Theta)$
\cite{Ouazi:2006}. Since ${\bar g_s}=g_s \langle S^z_{imp}(H,T)
\rangle$, it immediately follows that the droplet's scattering
strength and hence the resulting LDOS are expected to change with
$H$ and $T$. To demonstrate this effect, we present in
Fig.~\ref{fig3}(a) the total LDOS for several values of ${\bar g_s}$
representing different magnetic fields.
%
%  Fig.3
%
Consider, for example, that a given $H_0 \ll H^{ab}_{c2}$
corresponds to ${\bar g_s}=100$ meV. Increasing the magnetic field
to $2H_0$, (${\bar g_s}=200$ meV) or $4H_0$ (${\bar g_s}=400$ meV)
leads to a suppression of the LDOS in the droplet \cite{com5}. Since
the coupling between impurity and a charge mode is unaffected by the
magnetic field, the observation of a magnetic field dependent LDOS
as shown in Fig.~\ref{fig3}(a) is another direct signature of a
static spin droplet.

If both charge and a spin modes are present in the superconductor
and simultaneously couple to an impurity, we find that properties of
the resulting LDOS are predominantly determined by the ratio
$\alpha={\bar g}_c \eta_c/({\bar g}_s \eta_s)$, as shown in
Fig.~\ref{fig3}(b) for $\alpha=1/3$. As $\alpha$ changes from $0$ to
$\infty$, the LDOS changes continuously from being ``spin-like"
[Fig.\ref{fig1}(c)] to being ``charge-like" [Fig.\ref{fig1}(d)]. We
expect that the relative strength of the two modes can be determined
by studying the $H$-dependence of the LDOS and from the form of the
spin-resolved LDOS.

\begin{figure}[h]
 \epsfig{file=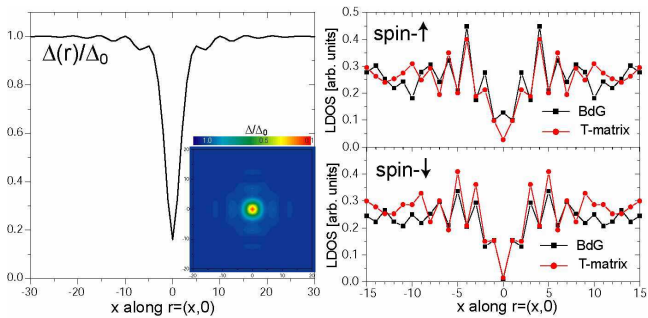,width=8.5cm}
 \caption{(color online) (a) SCOP along ${\bf r}=(x,0)$ for a spin droplet.
Inset: intensity plot of the SCOP. (b) Spin-$\uparrow$ and (c)
spin-$\downarrow$ LDOS of the ${\hat T}$-matrix (red line) and BdG
(black line) approaches at $\omega_{cp}=30$ meV along ${\bf
r}=(x,0)$.} \label{fig4}
\end{figure}

The pairbreaking nature of impurities in $d_{x^2-y^2}$-wave
superconductors leads to the suppression of the SCOP, an effect
which is not taken into account in the ${\hat T}$-matrix approach.
To study the relevance of this suppression, we compare the results
of the ${\hat T}$-matrix approach with those of the BdG formalism.
In Fig.~\ref{fig4}(a) we present the spatial form of the SCOP [see
Eq.(\ref{eq:BdGgap})] in a spin droplet for a system with $N=60
\times 60$ sites and the same parameters as above.
%
%  Fig.4
%
The SCOP is significantly suppressed only near the center of the
droplet, and quickly recovers its bulk value within a few lattice
spacings from the center. In order to ascertain how the SCOP's
suppression affects the LDOS, we compare in Fig.~\ref{fig4}(b)
[Fig.~\ref{fig4}(c)] the results of the ${\hat T}$-matrix and BdG
approaches for the spin-$\downarrow$ (spin-$\uparrow$) LDOS at
$\omega_{cp}$. The LDOS obtained from both approaches is in good
qualitative agreement and exhibits the same oscillations
characteristic of the antiferromagnetic nature of the droplet. The
quantitative differences are minimal: first, the LDOS of the BdG
approach undergoes a $\pi$-phase shift at $\Delta r \approx 10a_0$
from the center of the droplet, which is absent for the ${\hat
T}$-matrix results. Second, the spin-$\uparrow$ LDOS of both
approaches is out-of-phase at ${\bf r}=(0,0)$. A more detailed
analysis shows that the latter effect arises from a decrease of the
droplet's effective scattering strength due to the suppression of
the SCOP. Since similar good agreement is obtained for a charge
droplet, we conclude that the suppression of the SCOP has only minor
quantitative effects on the LDOS.

While we considered above a static spin droplet, recent studies
\cite{Morr:2003} suggest that our results remain valid even for
$H=0$ as long as the fluctuation time of the droplet satisfies
$\tau_S \gg 1/E_F$. If the spin excitations are sufficiently damped,
$\tau_S \rightarrow \infty$ and the droplet becomes static
\cite{Millis:2001}. One might therefore wonder whether signatures of
an impurity coupling to a quasi-static mode have already been
observed \cite{Hudson:2001}. The effects of collective modes on
quasi-particle interference patterns in the fourier-transformed LDOS
were discussed in Refs.\cite{DellAnna:2005,fluctmode}.

Finally, placing an impurity into the CuO$_2$-plane of the HTSC can
lead to static lattice distortions. Since such a ``phonon droplet"
is non-magnetic in nature, it gives rise to a scattering term in
Eq.(\ref{eq:ham_imp_int}) that is identical to that of the charge
droplet albeit with a potentially more complicated momentum
dependence of $g$. We therefore expect the qualitative effects of
static phonon and charge droplets to be quite similar.

In conclusion, we have shown that impurity induced static spin or
charge droplets exert qualitatively different effects on the LDOS of
a $d_{x^2-y^2}$ superconductor. We find that in a spin droplet, the
creation of a resonant impurity state is suppressed, the
spin-resolved LDOS exhibits complementary spatial patterns, and a
characteristic magnetic field dependence. Since these fingerprints
are absent in a charge droplet, our results demonstrate that
impurities can be used to identify the nature and relative strength
of collective modes not only in the HTSC, but in strongly correlated
electron systems in general.

We would like to thank J.C. Davis and D. Maslov for helpful
discussions and the Aspen Center for Physics for its hospitality.
D.K.M. acknowledges financial support by the Alexander von Humboldt
Foundation, the NSF under Grant No. DMR-0513415 and the U.S. DOE
under Award No. DE-FG02-05ER46225.


\begin{thebibliography}{99}
\bibitem{com3} Due to computational limitations, we set $\langle
s_e^z(\mathbf{r}) \rangle, \langle \delta n_e (\mathbf{r})
\rangle=0$ for $|\mathbf{r}-\mathbf{R}|>2\xi_{s,c}$. This leads to
only minor quantitative changes due to the exponential decay of the
droplet's amplitude.
\bibitem{com5} Note that for a decoupled impurity ($\eta_s=0$), the
impurity resonance shifts to lower frequencies with increasing $H$.
\end{thebibliography}
\end{document}